\documentclass[pre,amsmath,amssymb,twocolumn]{revtex4}

\usepackage{amsmath}    % need for subequations
\usepackage{amsfonts}
\usepackage{amssymb}
\usepackage{color,graphics}
\usepackage{graphicx}
\usepackage{dcolumn}
\usepackage{bm}

\begin{document}

\title{Robustness of the filamentation instability for asymmetric plasma shells collision in arbitrarily oriented magnetic field}

\author{A. Bret}
 \email{antoineclaude.bret@uclm.es}

\affiliation{ETSI Industriales, Universidad de Castilla-La Mancha, 13071 Ciudad Real, Spain}
\affiliation{Instituto de Investigaciones Energéticas y Aplicaciones Industriales, Campus Universitario de Ciudad Real, 13071 Ciudad Real, Spain}

\date{\today }

\begin{abstract}
The filamentation instability triggered when two counter streaming plasma shells overlap appears to  be the main mechanism by which collisionless shocks are generated. It has been known for long that a flow aligned magnetic field can completely suppress this instability. In a recent paper [PHYSICS OF PLASMAS 18, 080706 (2011)], it was demonstrated in two dimensions that for the case of two cold, symmetric, relativistically colliding shells, such cancellation cannot occur if the field is not perfectly aligned. Here, this result is extended to the case of two asymmetric shells. The filamentation instability appears therefore as an increasingly robust mechanism to generate shocks.
\end{abstract}

%\pacs{52.35.Qz, 52.35.Hr, 52.50.Gj, 52.57.Kk}

\maketitle

Collisionless shocks in plasmas have the capacity to accelerate particles up to high energies. This was first predicted \cite{Blandford78,Bell1978a,Bell1978b} before it was simulated \cite{Hededal2004,Spitkovsky2008a} and observed in space \cite{science_hecr}. As such, these structures are extremely promising candidates for the generation of high energy cosmic rays (HECR). Because the shock front, where particles are accelerated, is a region of strong magnetic turbulence, the resulting emission could, according to the so-called ``Fireball Scenario'' \cite{Piran2004}, equally explains the origin of gamma ray bursts (GRB).

While the birth of a fluid shock is well understood and can be described, for example, by the steepening of a large amplitude sound wave \cite{Zeldovich}, there is currently no established theory for the birth of a collisionless shock. Yet, both simulations \cite{Hededal2004,Spitkovsky2005,Spitkovsky2008a,Spitkovsky2008} and experiments have showed they can be generated by the encounter of two collisionless plasma shells \cite{CollLess_NJP_2011,Kuramitsu2011,Gregori2012}. In the absence of particle collisions, the shells simply start passing through each other. Soon, the overlapping region turns unstable to counter streaming plasmas instabilities, triggering the shock formation through a mechanism yet to be fully explained \cite{BretPoP2013}.

A detailed analysis of the unstable modes involved shows that in the relativistic regime, where GRB's and the most energetic HECR's could be produced, the dominant instability is the so-called ``filamentation'' instability \cite{BretPRL2008,BretPRE2010}. This instability amplifies perturbations with a wave vector normal to the flow. From the physical point of view, it is easily understood that whenever unbalanced counter streaming currents appear in the overlapping region as a result of a perturbation, both Lorentz' and Newton' laws will spontaneously amplify the unbalance.

The filamentation instability is therefore a key ingredient for the physics of collisionless shocks. Given the ubiquity of magnetic fields in astrophysical environments, their influence on the instability is worth investigating. For a cold system, it has been known for nearly 40 years that a flow aligned magnetic field can completely quench it \cite{Godfrey1975}. This result has been extended to hot systems since then \cite{Cary1981}. Recently, it was proved rigourously that still in the cold regime, a magnetic field can completely suppress the instability if and only if it is perfectly aligned with the flow \cite{BretAlvaroPoP}. In contrast, the instability growth rate remains finite, even in the large field limit. This late result is significant as it guarantees the occurrence of the instability, hence the shock formation, in realistic settings where the field could hardly be perfectly aligned \cite{BretPoPOblique,SironiApj}.

Yet, this result has been achieved only for the simple case of identical colliding plasma shells. The goal of this paper is to check this result in the more realistic case of asymmetric shells encounter. Due to the complexity of the calculations, the cold regime is still assumed for both shells. The main conclusion is that filamentation instability always persists for non-aligned magnetic field, even in the case of asymmetric counter streaming flows.

We therefore consider two 2D counter streaming electron/proton plasmas and ignore the protons' motion due to their highest inertia. The denser shell of initial density $n_{0p}$ comes from the right with initial velocity $\mathbf{v}_{0p}$ aligned with the $x$ axis. The dilute shell comes from the left with initial density $n_{0b}\ll n_{0p}$ and velocity $\mathbf{v}_{0b}$ aligned with the same $x$ axis. A static magnetic field $\mathbf{B}_0$ makes the angle $\theta$ with the axis $x$. Because each shell is both charge and current neutral, our system is initially charge and current neutral regardless of the reference frame. For convenience then, we choose to describe the process in a reference frame where $v_{0p}=(n_{0b}/n_{0p})v_{0b}$. This allows to bridge continuously from the symmetric case $\mathbf{v}_{0b}=-\mathbf{v}_{0p}$ to the diluted beam/plasma interaction regime with $n_{0b}\ll n_{0p}$ and $\mathbf{v}_{0p}\sim 0$ that we investigate.

For both shells we write the fluid conservation and momentum equations. For the rightward diluted shell, we have
\begin{eqnarray}\label{eq:b}
    \frac{\partial n_b}{\partial t}+\nabla\cdot(n_b\mathbf{v}_b)&=&0, \\
\frac{\partial \mathbf{p}_b}{\partial t}+(\mathbf{p}_b\cdot\nabla)\mathbf{p}_b&=&q\left[\mathbf{E}+\frac{\mathbf{v}_b\times(\mathbf{B}+\mathbf{B}_0)}{c}\right], \nonumber
\end{eqnarray}
with $\mathbf{p}_b=\gamma_b m \mathbf{v}_b$. For the leftward denser shell, we have
\begin{eqnarray}\label{eq:p}
    \frac{\partial n_p}{\partial t}+\nabla\cdot(n_p\mathbf{v}_p)&=&0,\\
    \frac{\partial \mathbf{v}_p}{\partial t}+(\mathbf{v}_p\cdot\nabla)\mathbf{v}_p&=&\frac{q}{m}\left[\mathbf{E}+\frac{\mathbf{v}_p\times(\mathbf{B}+\mathbf{B}_0)}{c}\right].\nonumber
\end{eqnarray}
By virtue of $n_{0b}\ll n_{0p}$ and $v_{0p}=(n_{0b}/n_{0p})v_{0b}$, this latter equation is non relativistic. In contrast, Eq. (\ref{eq:b}) is relativistic, i.e written in terms of the momentum $\mathbf{p}_b=\gamma_b m \mathbf{v}_b$, in order to account for rightward diluted shells with $v_{0b}$ close to $c$.

Note that while our system is initially charge and current neutral, it is not strictly in equilibrium because of the $y$ component of the magnetic field. On the one hand, the misaligned field affects the shells dynamic on a time scale given at least by,
\begin{equation}\label{eq:timeB}
  \frac{m c}{q B_0\sin\theta}\equiv \frac{1}{\omega_B\sin\theta}.
\end{equation}
On the other hand, the instability grows on a time scale $(\delta\omega_{pp})^{-1}$ where $\omega_{pp}$ is the electronic plasma frequency on the denser shell, and $\delta$ the growth rate in units of plasma frequency. Thus, the instability governs the early system dynamic provided,
\begin{equation}\label{eq:limit}
 \frac{1}{\omega_B\sin\theta} \gg \frac{1}{\delta\omega_{pp}}\Rightarrow \delta \gg \frac{\omega_B}{\omega_{pp}}\sin\theta.
\end{equation}

Eqs. (\ref{eq:b},\ref{eq:p}) are now linearized assuming small perturbations of all quantities proportional to $\exp(iky - i\omega t)$. The dielectric tensor has been computed analytically using a \emph{Mathematica} Notebook described elsewhere \cite{BretCPC}, and in terms of the reduced variables,
\begin{equation}\label{eq:variables}
  x = \frac{\omega}{\omega_{pp}}, ~~Z=\frac{k v_{0b}}{\omega_{pp}},~~
  \beta=\frac{v_{0b}}{c},~~\Omega_B = \frac{\omega_B}{\omega_{pp}},~~ \alpha=\frac{n_{0b}}{n_{0p}}.
\end{equation}

The dispersion equation $P(x)=0$ is a 24th degree polynomial in $x$ reported in the associated \emph{Mathematica} Notebook. The Notebook uses Eqs. (\ref{eq:b},\ref{eq:p}), together with Maxwell's equations and the conservation equations, all linearized and written in Fourier space, to derive the $3\times 3$ dielectric tensor of the system. The dispersion equation is then analytically obtained from the determinant of the tensor.

\begin{figure}
\begin{center}
 \includegraphics[width=0.45\textwidth]{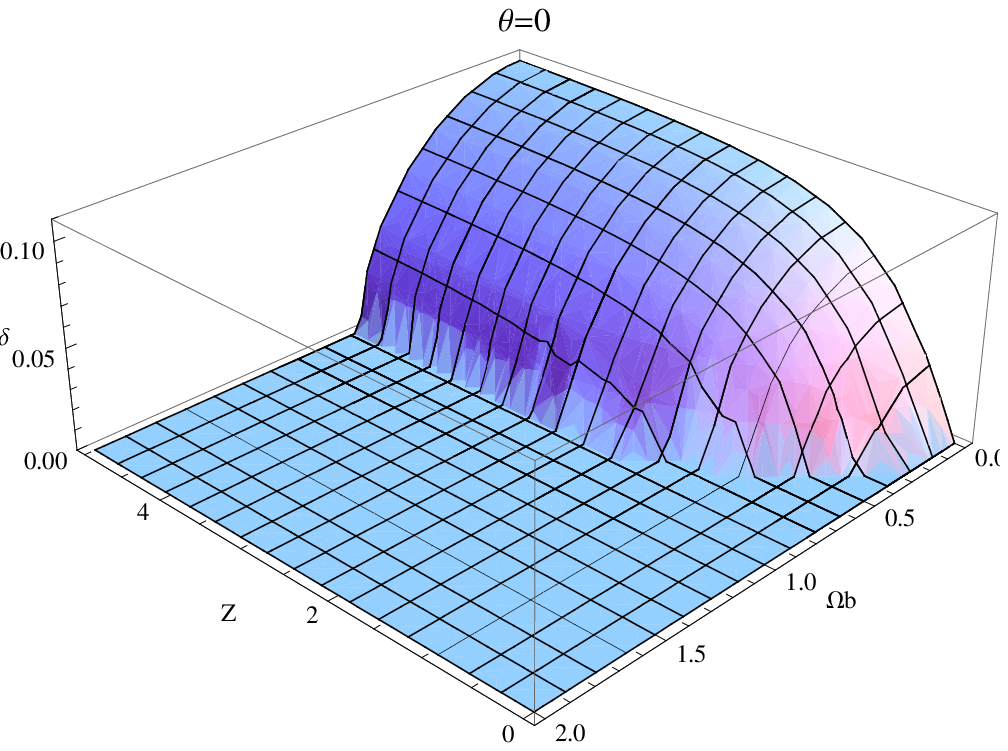}\\
 \includegraphics[width=0.45\textwidth]{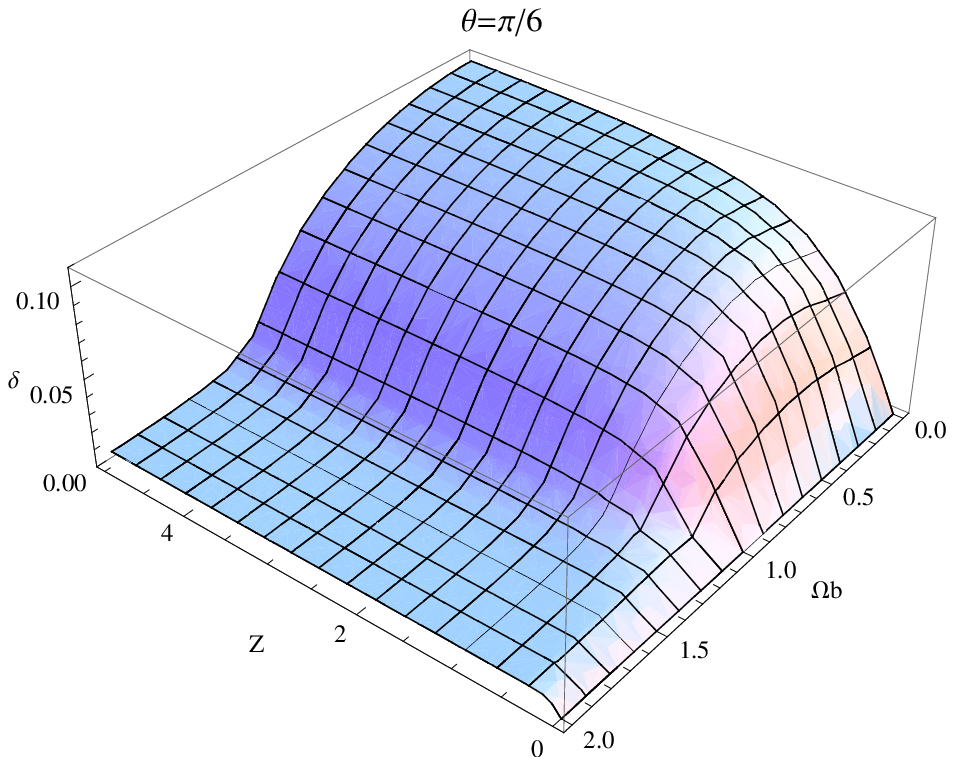}
\end{center}
\caption{(Color online) Growth rate $\delta$ in $\omega_{pp}$ units against the reduced wave vector $Z$ and the magnetic field parameter $\Omega_B$ for $\theta=0$ and $\pi/6$. Parameters are $\alpha=0.1$ and $\gamma_{0b}=10$.}
\label{fig:1}
\end{figure}

We start analysing the numerical resolution of the equation for various parameters. In this respect, figure \ref{fig:1} features the dimensionless growth rate $\delta$ against the reduced wave vector $Z$ and the magnetic field parameter $\Omega_B$, for $\theta=0$ and $\pi/6$. For $\theta=0$, we recover the known cancelation of the instability beyond a magnetic threshold given by $\Omega_B\sim \sqrt{\alpha/\gamma_{0b}}$ \cite{Godfrey1975}. For $\theta=\pi/6$, the pattern already highlighted in the symmetric case seems to remain valid: the instability displays a finite growth rate even at large $\Omega_B$'s.

\begin{figure}
\begin{center}
 \includegraphics[width=0.45\textwidth]{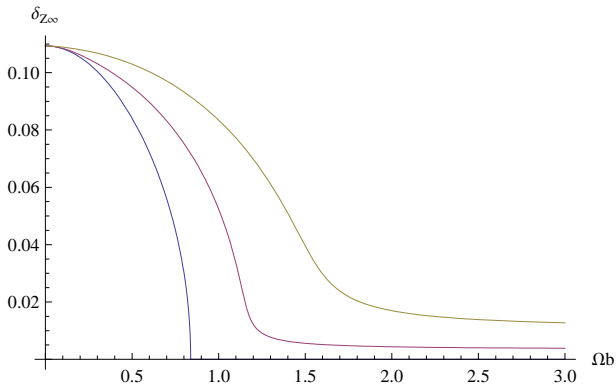}
\end{center}
\caption{(Color online) Growth rate $\delta_{Z\infty}$ in $\omega_{pp}$ in the limit $Z=\infty$ vs. the magnetic field parameter $\Omega_B$ for (low to high) $\theta=0$, $\pi/10$ and $\pi/4$. Same parameters as fig. \ref{fig:1}.}
\label{fig:2}
\end{figure}

In order to confirm this result, we start by exploiting the fact that the growth rate saturates for large $Z$'s. The dispersion equation in the limit of large $Z$, denoted $P_{Z\infty}(x)=0$, is simply obtained equaling to 0 the coefficient of the highest degree in $Z$ of the full dispersion equation $P(x)=0$. Figure \ref{fig:2} explains the numerical resolution of $P_{Z\infty}(x)=0$ for various values of $\theta$, confirming the result of fig. \ref{fig:1}.

Although the present theory is not valid in the large magnetic field limit by virtue of condition (\ref{eq:limit}), the saturation value of $\delta_{Z\infty}$ for large magnetic field is worth investigating, as it can yield a lower bound to the growth rate. The technique already applied to extract the asymptotic dispersion equation at $Z=\infty$ can be used here again. We simply extract the coefficient of the highest power (here, 4) of $\Omega_B$ in the polynomial $P_{Z\infty}(x)$. The resulting high-$Z$, high $\Omega_B$, dispersion equation reads,
\begin{eqnarray}\label{eq:GdB}
P_{Z\infty,\Omega_B\infty}(x)&=&\cos\theta^4 x^4 \gamma_{0b}^3-\sin\theta^4 (\alpha  \beta ^2+x^2 \gamma_{0b}-x^4 \gamma_{0b})\nonumber\\
&+&\cos\theta^2 \sin\theta^2 x^2 \gamma_{0b} \left(x^2+(x^2-1) \gamma_{0b}^2\right)\nonumber\\
&=&0.
\end{eqnarray}
In the limit $\alpha\ll 1$ where we work, the root of the equation yielding an exponentially growing mode reads,
\begin{equation}\label{eq:tauxOK}
\delta_{Z\infty,\Omega_B\infty}=\beta\sqrt{\frac{\alpha}{\gamma_{0b}}}\frac{1}{\sqrt{1+\gamma_{0b}^2\cot\theta^2}},
\end{equation}
which is just the result for the symmetric case corrected by $\sqrt{\alpha/2}$, square root of half the dilution factor.

Equation (\ref{eq:tauxOK}) displays an extremum for $\theta=\pi/2$ corresponding to the growth rate without field. Such result is consistent with previous ones \cite{BretPoPOblique, BretAlvaroPoP} and can be understood from the mechanism of the instability. In order to filament the beam, the instability has the particles move transversely to the flow. If $\mathbf{B}_0$ is also transverse to the flow, the Lorentz force associated with the displacement vanishes, and the instability is like field free. Noteworthily, this stands only in a two-dimensional geometry. In three dimensions, particles can move transversely and not along $\mathbf{B}_0$. Further studies will be required to check how filamentation behaves for wave vectors still normal to the flow, but outside the plan ($\mathbf{v}_{0b},\mathbf{B}_0$). At any rate, the present result guarantees the occurrence of the instability in 3D: calculation is 2D-like because it focuses on wave vectors in the plan  ($\mathbf{v}_{0b},\mathbf{B}_0$). But the formalism is 3D. Therefore, even if all other modes were stable in 3D, those investigated here would still be unstable.

We can therefore conclude that as long as the instability governs the early dynamic of the system, it cannot be completely canceled even for the case of two asymmetric shells collision. This result increases the robustness of the filamentation instability as a shock mediator.

Further works involve the investigation of kinetic effects. It turns out that such effects tend to reduce the instability, because velocity spread reduces the number of particles coupled to any given growing wave \cite{BretPoPReview}. Detailed calculations will thus be necessary to check wether the present result remains valid beyond the cold regime.

Thanks are due to Mark Dieckmann for enriching discussions.

%\bibliography{BibBret}

\end{document}